\documentclass[pra,showpacs,superscriptaddress,twocolumn,amsmath,amssymb]{revtex4}
\usepackage{graphicx}
\usepackage[pdfstartview=FitH,bookmarks=false]{hyperref}
\usepackage{bm}

\begin{document}

\title{Dynamics of Overhauser Field under nuclear spin diffusion in a quantum dot}
\author{Zhe-Xuan Gong} \affiliation{Department of Physics and MCTP, University of Michigan, Ann Arbor, Michigan 48109, USA}
\author{Zhang-qi Yin} \affiliation{Department of Physics and MCTP, University of Michigan, Ann Arbor, Michigan 48109, USA} \affiliation{Department of Applied Physics,
Xi'an Jiaotong University, Xi'an 710049, China}
\author{L.-M. Duan} \affiliation{Department of Physics and MCTP, University of Michigan, Ann Arbor, Michigan 48109, USA}
\date{\today}

\begin{abstract}
The coherence of electron spin can be significantly enhanced by locking the Overhauser field from nuclear spins using the nuclear spin preparation.
We propose a theoretical model to calculate the long time dynamics of the Overhauser field under intrinsic nuclear spin diffusion in a quantum dot.
We obtain a simplified diffusion equation that can be numerically solved and show quantitatively how the Knight shift and the electron-mediated
nuclear spin flip-flop affect the nuclear spin diffusion. The results explain several recent experimental observations, where the decay time of
Overhauser field is measured under different configurations, including variation of the external magnetic field, the electron spin configuration in a
double dot, and the initial nuclear spin polarization rate.
\end{abstract}

 \maketitle

\section{Introduction}

Electron spins in single quantum dots consist of one of the promising systems for realization of quantum computation \cite{LD98}. The spin state of a
single electron in a quantum dot can be coherently controlled either optically through fast laser pulses or electronically through tuning of gate
voltages \cite{LD98,ASL02,TED05,PJT05,XWS07}. In experiments, the coherence time of the electron spin is limited by its hyperfine coupling to the
nuclear spin environment in the host semiconductor material. The coupling causes spectral diffusion and gives a typical spin decoherence time
$T_{2}^{\ast }\sim 15ns$ for the electron spin qubit \cite{KLG02,HKP07}. This coherence time could be significantly prolonged with application of
spin echo or other dynamic decoupling techniques \cite{DD}. However, implementation of these techniques requires repeated applications of many laser
pulses. Each pulse inevitably induces some noise by itself, which limits the practical performance of the dynamic decoupling techniques to suppress
spin noise under real environments.

Another technique to increase the coherence time for the electron spin is through the dynamic nuclear-spin preparation (DNP)
\cite{NSP03,WS08,PTJ08,RTP08,XYS09,LHZ09,VNK09}, which prepares the nuclear spin environment into certain configurations. The nuclear spins in this
configuration collectively generate an effective magnetic field (the Overhauser field) on the electron spin with small fluctuation, significantly
reducing the spectral diffusion of the electron spin qubit. Although nuclear spins can be polarized through many methods such as optical pumping
\cite{NSP03}, substantial reduction in the fluctuation of the Overhauser field requires almost complete polarization of the nuclear spins
\cite{WS08}, which is hard to achieve experimentally. The recent experiments, however, demonstrate a surprising feedback mechanism which can lock the
Overhauser field to certain values without significant polarization of the nuclear spin environment \cite{RTP08,XYS09,LHZ09,VNK09}. The Overhauser
field generated from this locking mechanism has small fluctuation, which effectively increase the coherence time $T_{2}^{\ast }$ of the electron spin
qubit by up to two orders of magnitudes. The locking of the Overhauser field through the DNP process has been seen in a number of experimental
systems, under both optical and electronic gate control \cite{RTP08,XYS09,LHZ09,VNK09}.

To enjoy a longer spin coherence time, it is desirable to do gate experiments on the electron spins under the nuclear spin environment prepared with
the DNP,\ which produces a fixed Overhauser field. An important question is then how long this fixed Overhauser field can survive after the DNP
process and what facts determine/influence the relaxation time of the Overhauser field. Recent experiments have observed that the produced Overhauser
field can endure from a few seconds to a few minutes and the relaxation time of the Overhauser field depends on a variety of experimental parameters,
such as the applied magnetic field \cite{MBI07,Reily08,NCM09}, the electron spin configuration in double quantum dots \cite{Reily08}, and the DNP
pump time \cite{MTV08}.

In this paper, we develop a quantitative theory to calculate the relaxation time of the Overhauser field under the environment of quantum dots, and
provide an explanation to the experimental observations mentioned above in different setups and with control of different experimental parameters.
The relaxation of the Overhauser field is caused by nuclear spin diffusion, which has been well studied in bulk material \cite{LG67, Pagnet82}, where
diffusion is caused simply by the nuclear dipole-dipole interaction. In a quantum dot, however, the presence of the electron spin generates several
new effects. First, the electron spin can mediate the diffusion of the nuclear spins through a virtual hyperfine process. Second, the effective
magnetic field from the electron spin produces an inhomogeneous Knight shift on the nuclear spins through the hyperfine coupling which suppresses the
nuclear spin diffusion. The change of the nuclear spin diffusion coefficient by the Knight shift has been taken into account in Ref. \cite{DH05}, but
without consideration of the electron mediated nuclear spin diffusion. A recent work considers relaxation of the Overhauser field due to the electron
mediated nuclear spin diffusion, but without consideration of the direct nuclear dipole-dipole interaction \cite{KCL08}. In this case, the Overhauser
field can only decay by less than $1\%$, which suggests that it is necessary to include the dipole interaction in the long time dynamics. A
quantitative theory is still missing to our knowledge that includes a complete description of all the competing effects mentioned above. In this
work, we take into account all these diffusion mechanisms, and the resulting theory provides an explanation of the recent experimental observations
in Refs. \cite{MBI07,Reily08,NCM09,MTV08}.

The paper is arranged as follows: in Sec. II we give a formalism to describe relaxation of the Overhauser field that includes contributions from the
nuclear dipole-dipole interaction, the electron mediated nuclear spin diffusion, and the Knight shift. The effective nuclear spin diffusion equation
is solved numerically to determine the relaxation time of the Overhauser field. In Sec. III, we compare the theoretical calculations with the recent
experimental observations under control of different experimental parameters, and show that they are in qualitative or semi-quantitative agreement.
We summarize our results in Sec. IV with brief discussions.

\section{Decay of the Overhauser field through nuclear spin diffusion}

We assume that an external magnetic field $B_{0}$ much larger than the mean value and variance of the local Overhauser field generated by nuclei is
applied along the z-direction (perpendicular to the quantum dot layer). In this case, we can drop the nonsecular terms in the interaction Hamiltonian
\cite{Slichter63}. For simplification, we consider only one species of nuclei around the quantum dot electron. The total Hamiltonian for the electron
and nuclear spin system, including both the Fermi contact hyperfine interaction and nuclear dipole-dipole interaction, can be written as:

\begin{eqnarray}
H &=&H_{e}+H_{n}+H_{en}+H_{nn}, \\
H_{e} &=&-g_{e}\mu _{B}B_{0}S^{z}, \\
H_{n} &=&-g_{n}\mu _{N}B_{0}\sum_{i}{I_{i}^{z},} \\
H_{en} &=&\sum_{i}{A_{i}S^{z}I_{i}^{z}}+\sum_{i}{\frac{A_{i}}{2}(S^{+}I_{i}^{-}+S^{-}I_{i}^{+}),} \\
H_{nn} &=&\sum_{i\neq j}{2B_{ij}I_{i}^{z}I_{j}^{z}}-\sum_{i\neq j}{B_{ij}I_{i}^{+}I_{j}^{-},} \\
B_{ij} &=&\frac{\mu _{0}}{4\pi }(g_{n}\mu _{N})^{2}R_{ij}^{-3}(1-3\cos ^{2}\theta _{ij}),
\end{eqnarray}
where $A_{i}$ denotes the hyperfine coupling between the electron and nuclear spin at site $i$ with spatial coordinates $(x_{i},y_{i},z_{i})$.
$R_{ij}$ is the distance between two nuclei at site $i,j$. $\theta _{ij}$ is the angle between the line connecting sites $i,j$ and the z direction.

We note that for $B_{0}$ ranging from a few mT to a few T, the electron Zeeman splitting is on the order of $10^{-1}-10^{2}$GHz, while the average
hyperfine coupling in most quantum dot systems is on the order of MHz. Thus we can adiabatically eliminate the spin-flip terms in the hyperfine
interaction Hamiltonian and correspondingly modify the other terms in the Hamiltonian as \cite{KCL08}:
\begin{eqnarray}
H_{e} &=&-(g_{e}\mu _{B}B_{0}+\frac{\sum_{i}{A_{i}^{2}}}{4g_{e}\mu _{B}B_{0}})S^{z}, \\
H_{n} &=&\sum_{i}[-g_{n}\mu _{N}B_{0}+A_{i}(1-\frac{A_{i}}{4g_{e}\mu_{B}B_{0}})S^{z}]{I_{i}^{z},} \label{Hn}\\
H_{nn} &=&\sum_{i\neq j}{2B_{ij}I_{i}^{z}I_{j}^{z}}-\sum_{i\neq j}{(B_{ij}+\frac{A_{i}A_{j}{S^{z}}}{2g_{e}\mu _{B}B_{0}})I_{i}^{+}I_{j}^{-},}
\end{eqnarray}
where we have introduced an electron-mediated nuclear flip-flop term in $H_{nn}$. Since we are interested in the long time dynamics of nuclear spins,
we can completely eliminate the electron from the Hamiltonian by replacing the constant operator $S^{z}$ with its expectation value. We find that
using $S^{z}=1/2$ or $S^{z}=-1/2$ will yield almost the same result in the following calculations. Therefore we can set $S^{z}=1/2$ for simplicity
and arrive at the following effective Hamiltonian (neglecting constant terms):
\begin{eqnarray}
H &=&H_{0}+H_{1},  \label{H1} \\
H_{0} &\approx &\sum_{i}(-g_{n}\mu _{N}B_{0}+A_{i}/2){I_{i}^{z}}+\sum_{i\neq j}{2B_{ij}I_{i}^{z}I_{j}^{z},}  \label{H2} \\
H_{1} &=&-\sum_{i\neq j}{(B_{ij}+\frac{A_{i}A_{j}}{4g_{e}\mu _{B}B_{0}})I_{i}^{+}I_{j}^{-}.}  \label{H3}
\end{eqnarray}
Here, the term proportional to $A_{i}$ in $H_{0}$ is the Knight shift term. For this Knight shift, we have neglected the small term proportional to
$A_{i}^{2}$ in Eq. (\ref{Hn}) as it is dominated by the $A_{i}$ term.

The expectation value for $z$ component of the nuclear spin at site $k$ will evolve according to the Schrodinger equation:
\begin{equation}
\frac{\partial \langle I_{k}^{z}\rangle }{\partial t}=\frac{i}{\hbar }Tr\{\rho (t)[H_{1},I_{k}^{z}]\},
\end{equation}
where $\rho (t)$ is the nuclear spin density matrix at time $t$, which can be calculated by switching to the interaction picture:
\begin{equation}
\tilde{\rho}(t)=\rho (0)+\frac{i}{\hbar }\int_{0}^{t}{[\tilde{\rho}(t^{\prime }),\tilde{H}_{1}(t^{\prime })]dt^{\prime },}
\end{equation}
with $\tilde{H}_{1}(t)=exp(iH_{0}t/\hbar )H_{1}exp(-iH_{0}t/\hbar )$. Further calculation yields \cite{LG67}:
\begin{eqnarray}
&&\frac{\partial \langle I_{k}^{z}\rangle }{\partial t}=\frac{i}{\hbar }Tr\{\rho (0)[\tilde{H}_{1}(t),I_{k}^{z}]\}  \nonumber  \label{Ikz} \\
&+&\left( \frac{i}{\hbar }\right) ^{2}\int_{0}^{t}{Tr\{\rho (t-t^{\prime })[H_{1},[\tilde{H}_{1}(t^{\prime }),I_{k}^{z}]]\}dt^{\prime }}
\end{eqnarray}

We assume the nuclear spin (with spin-$I$) density matrix in a product state of the following form:
\begin{equation}
\rho (t)=\bigotimes_{k}\rho _{k}(t)\quad \rho _{k}(t)=\frac{1}{2I+1}+\frac{\langle I_{k}^{z}(t)\rangle }{Tr\{(I_{k}^{z})^{2}\}}I_{k}^{z}. \label{rho}
\end{equation}
Such an approximation is valid for the initial state of nuclear spin when correlations and transverse coherence are negligible in its equilibrium
configuration. Off-diagonal terms may appear in $\rho (t)$ for $t>0$, but it can be shown that they only have a minor contribution to the evolution
of $\langle I_{k}^{z}(t)\rangle $ compared to the diagonal part expressed in Eq. (\ref{rho}) (see Ref. \cite{LG67}).

By using the explicit form of the Hamiltonian [Eq. (\ref{H1}-\ref{H3})] and density matrix [Eq. (\ref{rho})], we can reduce Eq. (\ref{Ikz}) to:
\begin{eqnarray}
\frac{\partial \langle I_{k}^{z}\rangle }{\partial t} &=&\sum_{i\neq k}W_{ik}(\langle I_{i}^{z}(t)\rangle -\langle I_{k}^{z}(t)\rangle )
\label{W} \\
W_{ki} &=&\frac{1}{Tr\{(I_{k}^{z})^{2}\}}\int_{0}^{t}{Tr\{[\tilde{H}_{1}(t),I_{k}^{z}][\tilde{H}_{1}(t-t^{\prime }),I_{i}^{z}]\}dt^{\prime }}
\nonumber
\end{eqnarray}
where $W_{ki}$ has a clear physical meaning as the flip-flop rate between nuclear spins at site $i$ and $k$.

For a 2D InAs/GaAs quantum dot, we take As nuclei ($I=3/2$) as an example for further calculation. The parameter $W_{ki}$ can be analytically
calculated when approximating the integration upper limit in the above expression for $W_{ki}$ from $t$ to infinity \cite{DH05}.
\begin{eqnarray}
W_{ik} &=&\frac{17\sqrt{2\pi }}{5}C_{ik}^{2}(A_{ik}^{2}+g_{ik})^{-1/2}
\nonumber \\
&+&\frac{12\sqrt{2\pi }}{5}C_{ik}^{2}(A_{ik}^{2}+64C_{ik}^{2}+g_{ik})^{-1/2}
\nonumber \\
&+&\frac{9\sqrt{2\pi }}{10}C_{ik}^{2}(A_{ik}^{2}+256C_{ik}^{2}+g_{ik})^{-1/2}, \\
A_{ik} &=&A_{i}-A_{k}, \\
C_{ik} &=&B_{ik}+\frac{A_{i}A_{k}}{4g_{e}\mu _{B}B_{0}}, \\
g_{ik} &=&80\sum_{j\neq i,k}(C_{ij}-C_{kj})^{2}.
\end{eqnarray}
The hyperfine coupling rate $A_{i}$ is proportional to square of the electron wave function in a quantum dot. And in the following
calculation we assume the dot potential is like a square well in the $z$-direction and the electron wave function takes a Gaussian shape in the $x,y$-plane. The hyperfine coupling rate $A_{i}$ can then be written as $%
A_{i}=A_{0}\cos ^{2}(\pi z_{i}/z_{0})\exp [-(x_{i}^{2}+y_{i}^{2})/l_{0}^{2}], $ where $(x_{i},y_{i},z_{i})$ are spatial coordinate of the site $i$.
$l_{0}$ and $z_{0}$ are, respectively, the Fock-Darwin radius and thickness of the quantum dot, and $A_{0}$ is the hyperfine coupling for the nuclear
spin at the origin where the electron locates.

We assume the nuclei spins follow a diffusion process, where the flip-flop rate for two distant sites is negligible. This approximation is justified
by the fact that the coefficient $C_{ik}$ generally decays to zero fast as $R_{ik}$ increases. If we treat $\langle I_{k}^{z}(t)\rangle $ as a
continuous function of spatial variable $x^{\alpha }$ ($\alpha =x,y,z$). We can then carry out a Taylor expansion of $\langle I^{z}(t)\rangle $ for
site $i$ around site $k$:
\begin{eqnarray}
\langle I_{i}^{z}(t)\rangle &\approx &\langle I_{k}^{z}(t)\rangle +\frac{\partial \langle I_{k}^{z}(t)\rangle }{\partial x^{\alpha }}(x_{k}^{\alpha}-x_{i}^{\alpha })  \nonumber \\
&+&\frac{1}{2}\frac{\partial ^{2}\langle I_{k}^{z}(t)\rangle }{\partial x^{\alpha }\partial x^{\beta }}(x_{k}^{\alpha }-x_{i}^{\alpha
})(x_{k}^{\beta}-x_{i}^{\beta })+\cdots
\end{eqnarray}
where Einstein's summation convention is implied for spatial index $\alpha ,\beta $. Substituting this into Eq. (\ref{W}) and noting that the summation
of the first order derivative term over all sites vanishes due to the lattice symmetry \cite{note}, we have:
\begin{equation}
\frac{\partial \langle I_{k}^{z}\rangle }{\partial t}\approx \lbrack \sum_{i\approx k}\frac{1}{2}W_{ik}(x_{k}^{\alpha }-x_{i}^{\alpha })(x_{k}^{\beta
}-x_{i}^{\beta })]\frac{\partial ^{2}\langle I_{k}^{z}(t)\rangle }{\partial x^{\alpha }\partial x^{\beta }}  \label{Dif1}
\end{equation}
The $\sum_{i\approx k}$ notation above means summarization over the sites near $k$. Define coefficient $D^{\alpha \beta }=\sum_{i\approx
k}W_{ik}(x_{k}^{\alpha }-x_{i}^{\alpha })(x_{k}^{\beta }-x_{i}^{\beta })/2$ and similarly note that for $\alpha \neq \beta $ the summation over all
sites vanishes, we have:
\begin{equation}
\frac{\partial \langle I_{k}^{z}\rangle }{\partial t}=(D^{xx}\frac{\partial ^{2}}{\partial x^{2}}+D^{yy}\frac{\partial ^{2}}{\partial
y^{2}}+D^{zz}\frac{\partial ^{2}}{\partial z^{2}})\langle I_{k}^{z}(t)\rangle  \label{Dif2}
\end{equation}

Eq. (\ref{Dif2}) is a 3D anisotropic diffusion equation with spatially varying diffusion coefficients (as $A_{ik}$, $B_{ik}$, $W_{ik}$ all depend on the
spatial coordinates), which is not easy to solve. To further simplify it, we note that to obtain the major feature for the full time dynamics of the
Overhauser field $\langle h_{z}(t)\rangle =\sum_{k}A_{k}\langle I_{k}^{z}(t)\rangle $, it is reasonable to first ignore the diffusion in the $z$
direction as the quantum dot layer is usually a few nm thick and chemical or structural mismatch in adjacent layers may strongly suppress diffusion
in the $z$ direction \cite{NCM09}. In addition, from symmetry in the 2D $x,y$-plane, we expect to have $D^{xx}\approx D^{yy}$ and thus define an
average 2D diffusion coefficient $D(x,y)=\sum_{i\approx k}W_{ik}[(x_{k}-x_{i})^{2}+(y_{k}-y_{i})^{2}]/4$. Now we have a simplified 2D diffusion
equation:
\begin{equation}
\frac{\partial \langle I_{k}^{z}\rangle }{\partial t}=D(x,y)(\frac{\partial ^{2}}{\partial x^{2}}+\frac{\partial ^{2}}{\partial y^{2}})\langle
I_{k}^{z}(t)\rangle .  \label{Dif3}
\end{equation}

The above partial differential equation can be effective solved using the finite element method by coarse graining a large number of nuclear spin
sites to a small number of mesh nodes. But before solving Eq. (\ref{Dif3}), we would like to have some discussion about the diffusion coefficient
$D(x,y) $. For $x,y\gg l_{0}$, the role of electron can be neglected and a numerical calculation of the above diffusion coefficient gives a uniform
background value $D\approx 7$ $nm^{2}/s$, which is consistent with the previous theoretical and experimental study of the diffusion coefficient in
the bulk material \cite{LG67,DH05,Pagnet82}. In our calculation, we set the quantum dot parameters as $l_{0}=30$ $nm,$ $z_{0}=10$ $nm,$ $A_{0}=1$
$\mu eV\approx 1.5$ $MHz,$ $\sum_{k}A_{k}\approx 80$ $\mu eV$, the lattice constant $a_{0}=0.563$ $nm$, and the number of nuclei $N\approx 9\times
10^{5}$, according to the typical experimental values \cite%
{MBI07,Reily08,NCM09}.

Within the range of Fock-Darwin radius $l_{0}$, however, the presence of the quantum dot electron will change the diffusion coefficient through two
competing mechanisms: on the one hand, the presence of the confined electron generates an inhomogeneous Knight shift \cite{KLG02}, which lifts the
degeneracy of the nuclear Zeeman energy for different nuclei and prevents the spin flip-flop; on the other hand, electron mediated nuclear spin
flip-flop enhances the nuclear spin diffusion from the center to the edge of the dot.

\begin{figure}
\includegraphics[width=0.5\textwidth]{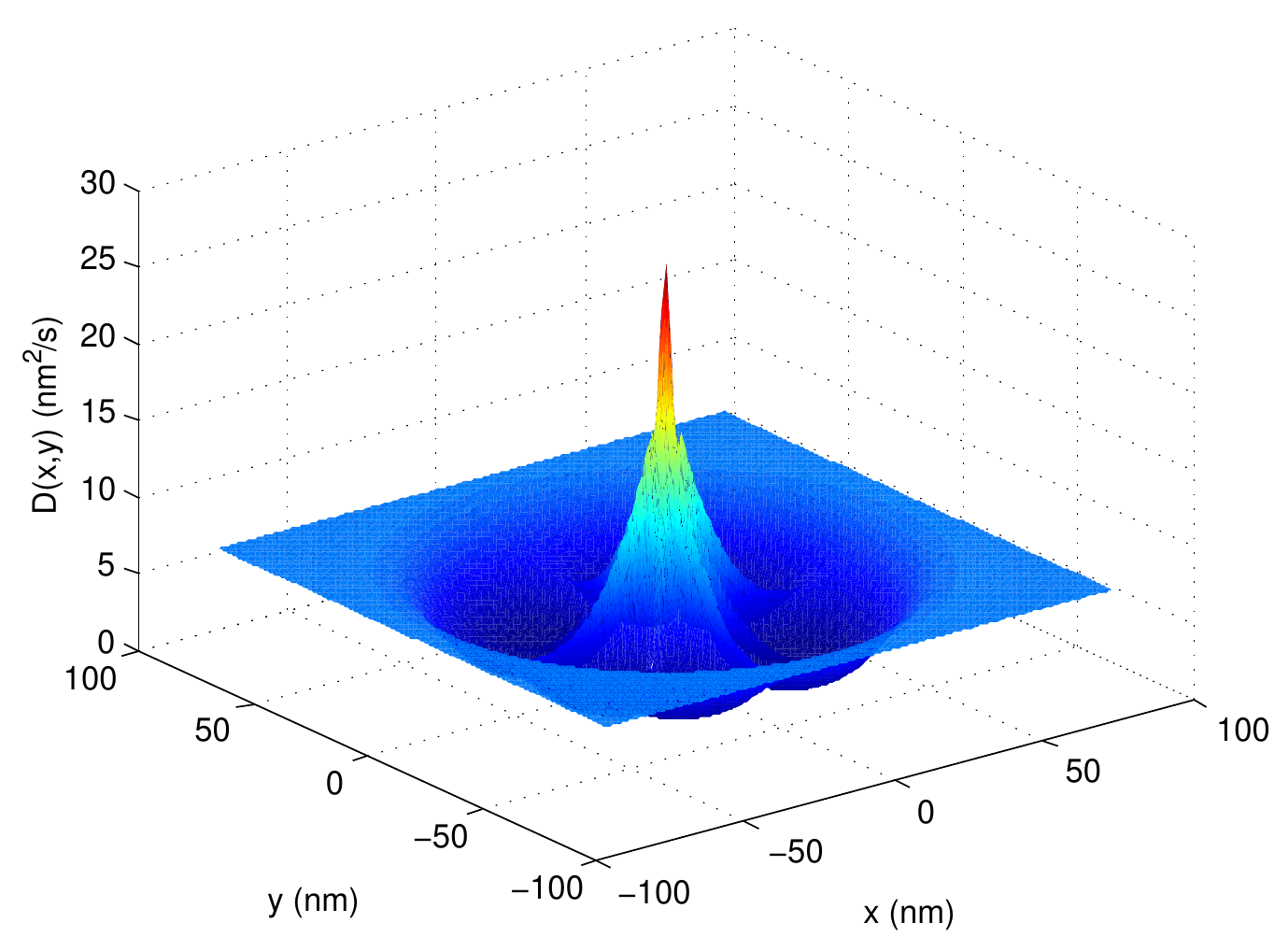}
\caption{Diffusion Coefficient D(x,y) under $B_0=0.2T$. The narrow high peak at the center of the dot is due to electron-mediated nuclear spin
flip-flop, and the wide dip is due to inhomogeneous Knight shift.}\label{D1}
\includegraphics[width=0.5\textwidth]{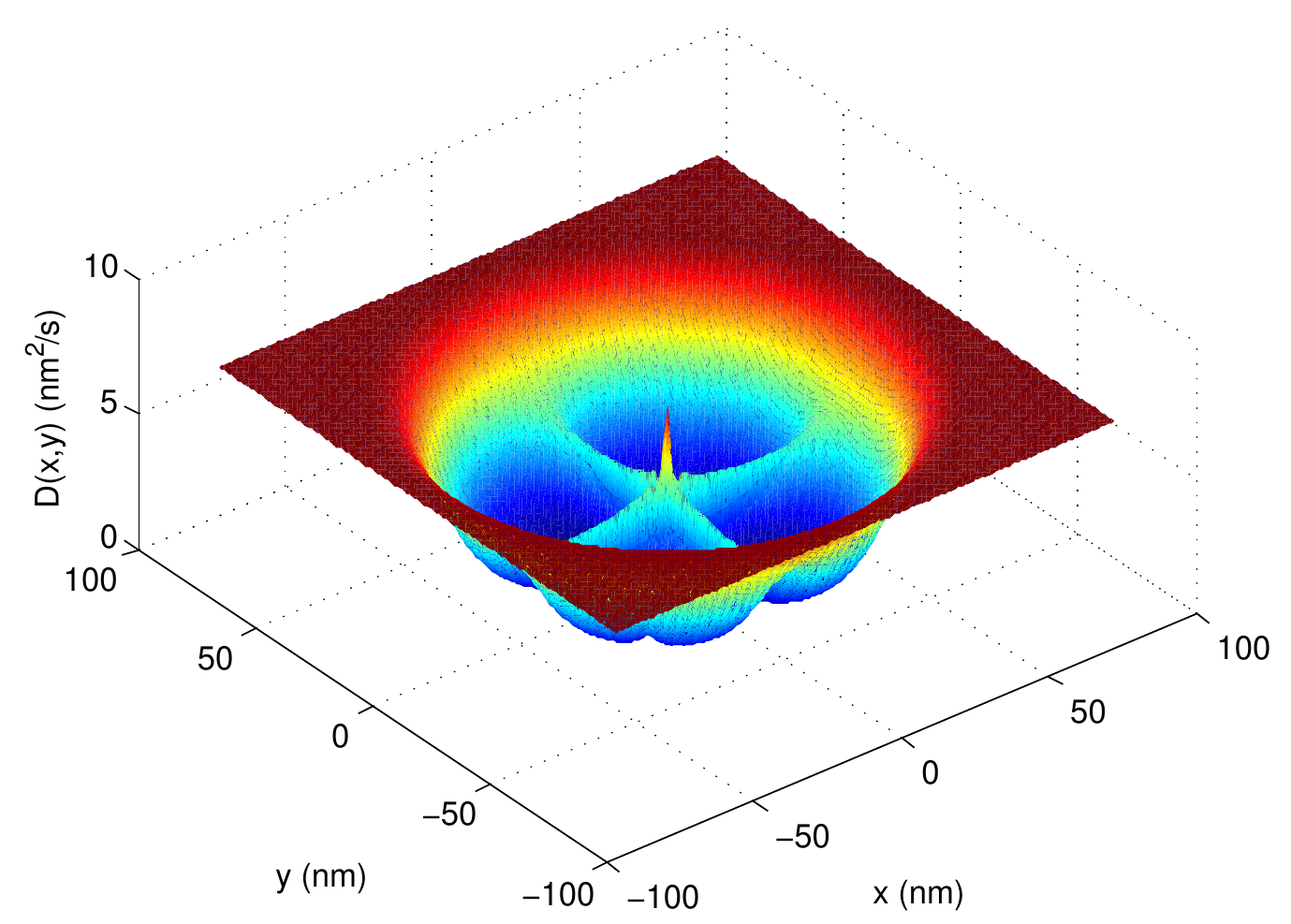}
\caption{Diffusion Coefficient D(x,y) under $B_0=2T$. The inhomogeneous Knight shift dominates in this case, so diffusion is generally suppressed
within the Fock-Darwin radius.}\label{D2}
\end{figure}

Our numerical simulation shows that whether one mechanism dominates the other is largely determined by the external magnetic field $B_{0}$.
Fig. \ref{D1}-\ref{D2} show the diffusion coefficient $D(x,y)$ under $B_{0}=0.2T$ and $B_{0}=2T$. We can see that the electron mediated flip-flop
greatly enhances the nuclear spin diffusion near the center of the dot under a small magnetic field (with a sharp peak in $D(x,y)$), while for a
large magnetic field, this effect is negligible compared to the Knight shift which suppresses the nuclear spin diffusion (with a wide dip in
$D(x,y)$). The difference can be easily explained from the effective Hamiltonian [Eq. (\ref{H1}-\ref{H3})]: the electron mediated flip-flop term is
inversely proportional to $B_{0}$ while the Knight shift term is independent of $B_{0}$. We note that the reason why we have a narrower peak than the
dip is due to the fact that the Knight shift term is proportional to the hyperfine coupling (with a Gaussian distribution) while the Electron
mediated flip-flop term is proportional to the product of two nuclei's hyperfine coupling rates. We also note that the 2D diffusion coefficient
$D(x,y)$ in Fig. \ref{D1}-\ref{D2} does not have azimuthal symmetry since in our calculation we assume the nuclear spins in a square lattice which
has no azimuthal symmetry.

\section{Comparison with experiments}

To compare with experiments, we numerically solve the diffusion equation [Eq. (\ref{Dif3})] under certain initial and boundary conditions. For the
initial condition, the nuclear spins are partially polarized through the DNP process from the hyperfine interaction with the electron spin
\cite{RTP08,XYS09,LHZ09,VNK09}. It is reasonable to expect that right after the DNP process, the polarization distribution $\langle I_{k}^{z}\rangle
$ is proportional to the hyperfine interaction rate. So in the following calculation, we assume $\langle I_{k}^{z}\rangle \propto A_{k}\propto \exp
[-(x_{i}^{2}+y_{i}^{2})/l_{0}^{2}] $ at $t=0$\ for solving the diffusion equation [Eq. (\ref{Dif3})]. We can take the natural boundary condition where
$\langle I_{k}^{z}\rangle $ approaches zero when the radius goes to infinity. However, in numerical calculation, we have to take a finite radius. To
make the spin diffusion possible, this finite radius has to be significantly larger than the radius of the size $l_{0}$ of the initial electron wave
packet. In the calculation, we typically take the integration size about $300 $ $nm$ ($10$ times of $l_{0}$) so that the number of total nuclear
spins in this area is about $100$ times the number of initially partially polarized nuclear spins covered by the electron wave packet. With this
choice, we expect the cutoff error should be small (at a percent level).

\begin{figure}
\includegraphics[width=0.5\textwidth]{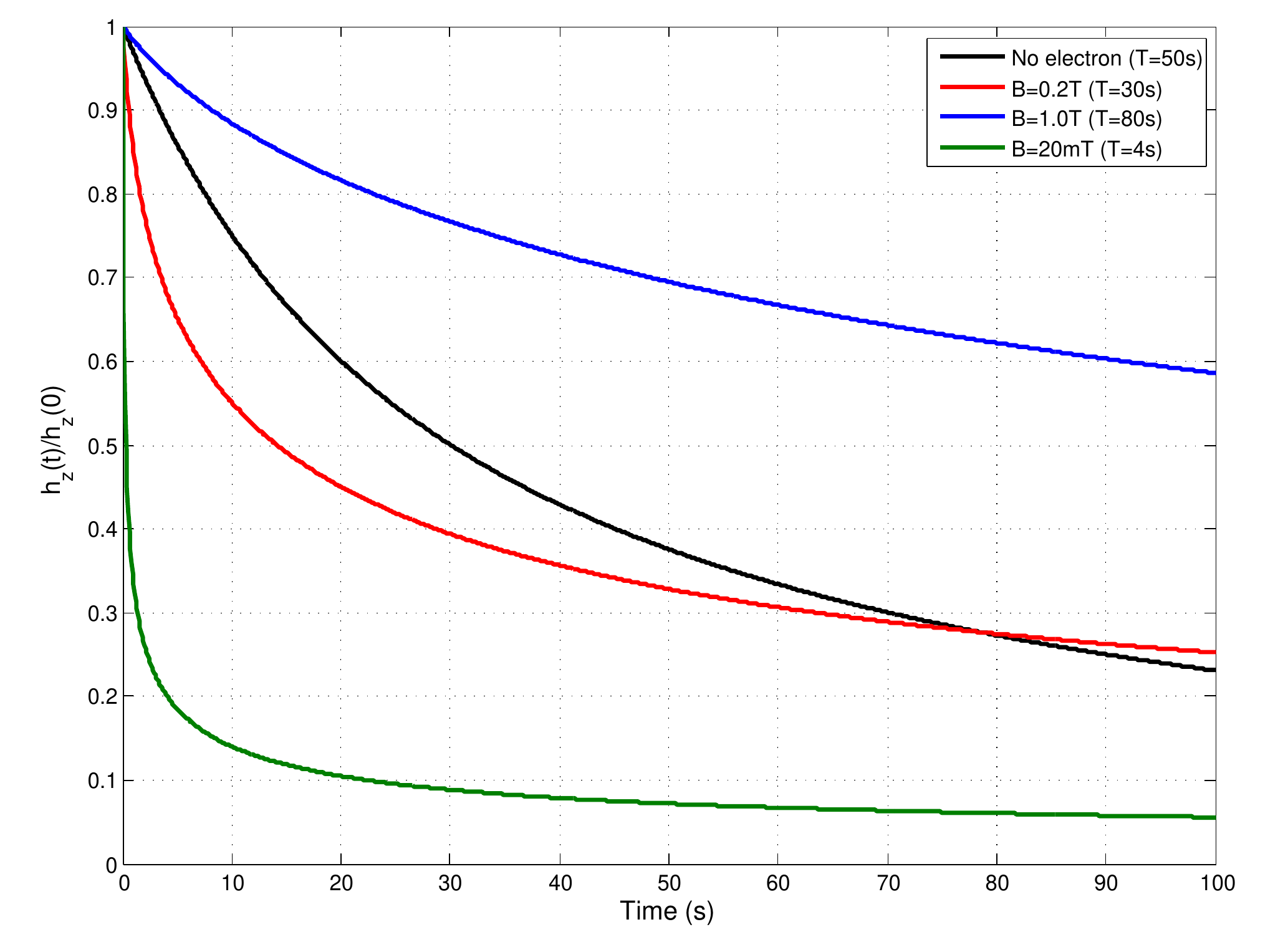}
\caption{Decay of the Overhauser field under various small magnetic fields. The solid line refers to the case with electron staying in the (2,0)
singlet state where electron plays no role in nuclear spin diffusion.} \label{decay1}
\end{figure}
\begin{figure}
\includegraphics[width=0.5\textwidth]{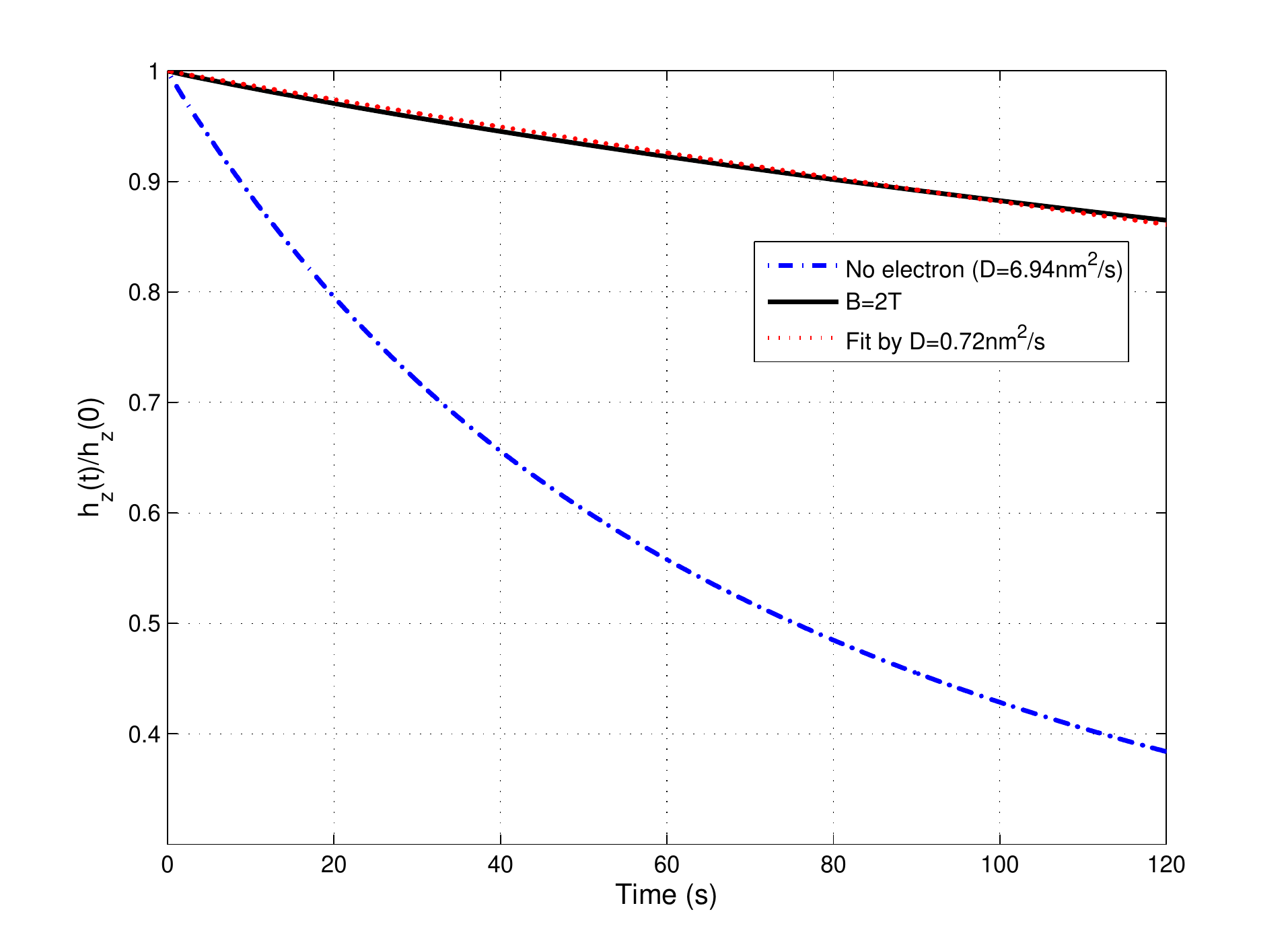}
\caption{Decay of the Overhauser field under a strong magnetic field. The solid line shows the dynamics of the Overhauser field under $B=2T$ and the
dotted line is a fit by using a constant diffusion coefficient.} \label{decay2}
\end{figure}

First, to compare with the experiments in Ref. \cite{Reily08}, we calculate the relaxation of the Overhauser field $h^{z}(t)=\sum_{k}A_{k}\langle
I_{k}^{z}(t)\rangle $ under different electron states and different values of the external magnetic field $B_{0}$, and the result is shown in
Fig. \ref{decay1}. We note that for the double quantum dot system in Ref. \cite{Reily08}, if the electron stays in the (2,0) singlet state, the
electron spin has $S_{z}\equiv 0$, with basically no influence on the nuclear spin diffusion. In this case, the nuclear spin diffusion is governed by
the intrinsic nuclear dipole-dipole interaction. However, for the electron in the (1,1) state, with the magnetic field in the range of tens of mT as
in this experiment, the electron mediated spin diffusion dominates the Knight shift and it accelerates the nuclear spin relaxation. One can see two
effects from Fig. \ref{decay1} with the magnetic field $B_{0}=10$ mT or $20$ mT: (1) electrons in the (1,1) state will speed up the decay of the
Overhauser field compared to electrons in the (2,0) state; (2) a smaller magnetic field gives a faster decay of the Overhauser field. Both of these
effects agree with the experimental observations in Ref. \cite{Reily08}, and the decay time scale is also consistent with what one measures from the
experiments in term of the order of magnitude.

With a much larger magnetic field (say, $B_{0}=2T$, as in experiments in Ref. \cite{NCM09}), the electron mediated nuclear spin diffusion is
suppressed, and the Knight shift plays a more important role. The Knight shift suppresses the nuclear spin diffusion, and can prolong the relaxation
time of the Overhauser field to make it significantly larger than the relaxation time in the bulk material. Fig. \ref{decay2} shows decay of the
Overhauser field in this case, and we can fit the curve with an effective constant diffusion coefficient with its value about $D_{eff}\approx
0.7nm^{2}/s$. Compared with the diffusion coefficient in the bulk material ($D\approx 7nm^{2}/s$), the effective diffusion here is suppressed by a
factor of $10$ under a strong magnetic field $B_{0}$. Experiments done in Ref. \cite{NCM09} yield an effective diffusion coefficient $50$ times
smaller than the value in the bulk system. While the suppression there could have contribution from other factors, such as inhomogeneity of lattice
constant there, we believe that the Knight-shift induced suppression plays a large role in this experiment under a strong magnetic field.

\begin{figure}
\includegraphics[width=0.5\textwidth]{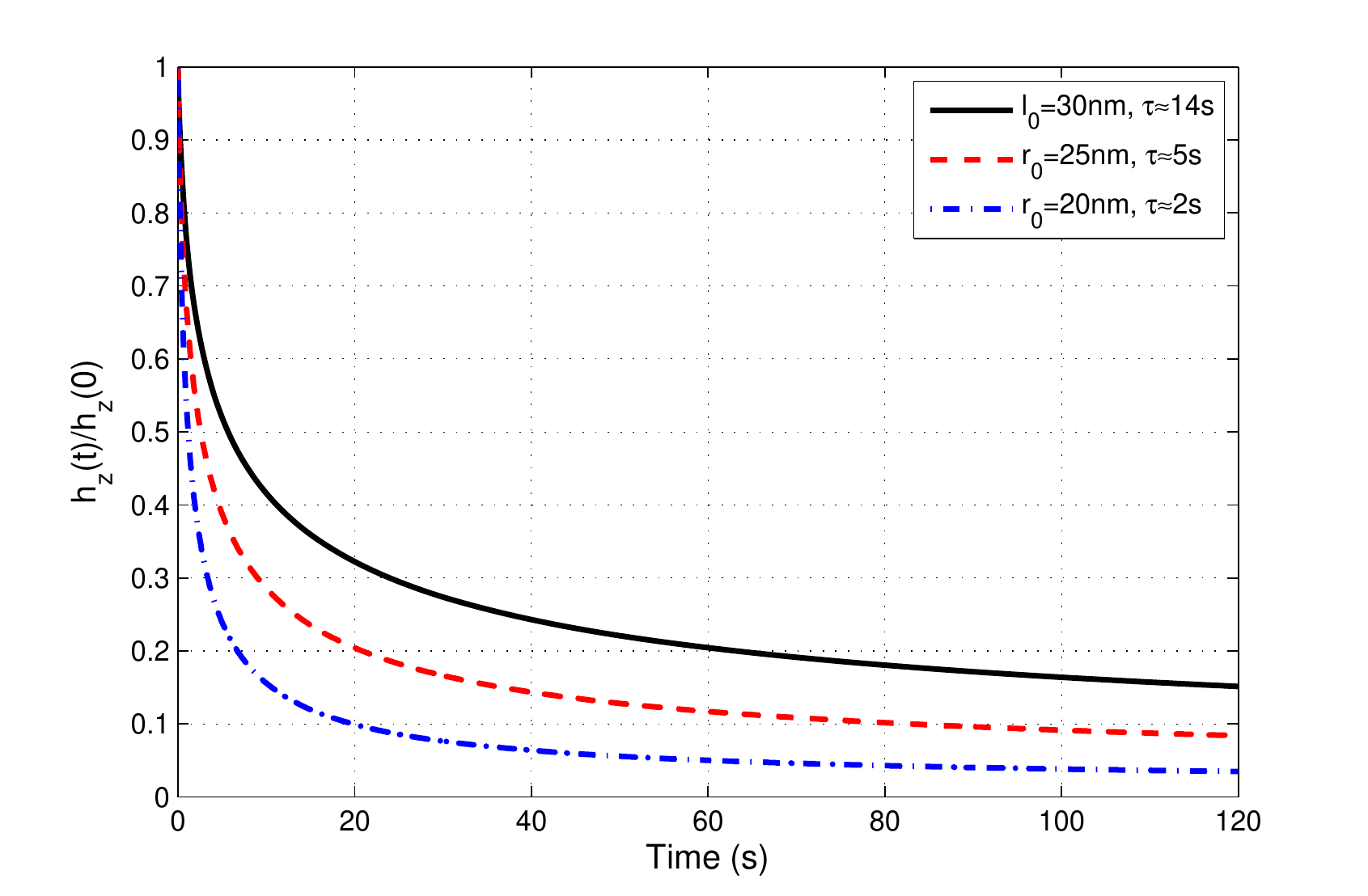}
\caption{Overhauser Decay of the Overhauser field under various initial distributions of the nuclear polarization. The solid line refers to the case
with a long DNP pump time that gives a Gaussian distribution with the size characterized by the Fock-Darwin radius $l_{0}$. Other lines correspond to
narrower polarization distribution characterized by a Gaussian with its size $r_{0}<l_{0}$.} \label{decay3}
\end{figure}

The experiment in Ref. \cite{MBI07} studies relaxation of the Overhauser field under different pumping time for the DNP process. With a shorter DNP
pumping time, the nuclear spin polarization may have a narrower distribution in space \cite{MBI07}. Although we do not know the exact distribution of
the nuclear spin polarization from a short DNP pump, to model this effect qualitatively, we simply assume that this distribution $\langle
I_{k}^{z}\rangle $ is still a Gaussian but with its radius $r_{0}<l_{0}$. Taking this $\langle I_{k}^{z}\rangle $ as the initial condition, we can
calculate relaxation of the corresponding Overhauser field from the diffusion equation [Eq. (\ref{Dif3})], and the result is shown in
Fig. \ref{decay3}. The result indicates that a narrower distribution of initial nuclear spin polarization leads to a faster decay of the Overhauser
field, which is consistent with the experimental result in Ref. \cite{MBI07}. This effect can be explained by noting that the diffusion is much
stronger near the center of the dot due to the electron mediated diffusion peak (see Fig. \ref{D1}), which results in a shorter relaxation time of the
Overhauser field if the initial polarization is more concentrated near the dot center.

\section{Summary and discussion}

In summary, we have established an effective method for calculating the long time dynamics of the Overhauser field under nuclear spin diffusion and
shown that the confined electron in a quantum dot can both enhance decay of the Overhauser field by mediating nuclear spin flip-flop and suppress
nuclear spin diffusion via inhomogeneous Knight shift. Which effect dominates depends critically on the magnitude of the external magnetic field. We
simulate the relaxation of Overhauser field under different electron spin configuration, external magnetic field, and initial nuclear polarization
distribution, and the results agree qualitatively with a series of recent experimental observations.

For the purpose of maintaining the Overhauser field generated by the DNP process as long as possible, applying a large magnetic field turns out to be
the most effective way since a large magnetic field suppresses the electron mediated nuclear spin diffusion. The intrinsic nuclear spin diffusion
from the nuclear dipole-dipole interaction is also suppressed by the inhomogeneous Knight shift. These two kinds of suppression, combined together,
leads to a long relaxation time of the Overhauser field.

For calculations in this paper, we focus on the relaxation dynamics of the expectation value of the Overhauser field, since this is the quantity that
has been measured in several recent experiments. Similar methods could also apply to calculation of relaxation of the variance of the Overhauser
field, and in term of the time scale, the relaxation time for variance is basically the same as the relaxation time for the expectation value.

During the DNP process, nuclear spin diffusion also takes place. The final distribution of the nuclear spin polarization and the limit on its
fluctuation may depend on a balance between the DNP pumping cycle and the continuous nuclear spin diffusion process
\cite{RTP08,XYS09,LHZ09,VNK09,Reily08}. To understand this balance, we need a more detailed understanding of the dynamics for the DNP process and its
dependence on various experimental control parameters. This is an interesting topic for further investigation.

\acknowledgements

We thank Duncan Steel for helpful discussions. This work is supported by the ARO MURI program, the IARPA grants, the DARPA OLE Program, and the AFOSR
MURI program.


\begin{thebibliography}{99}
\bibitem{note} {The hyperfine coupling rate $A_{i}$, although inhomogeneous over a larger scale, only changes slightly from one nuclear spin to its nearby spins, so in summation over the nearby lattice sites, we still have an approximate lattice symmetry for the jump rate $W_{ik}$.}
\end{thebibliography}
\end{document}